# A Full-Potential-Linearized-Augmented-Plane-Wave Electronic Structure Study of δ - Plutonium and the (001) Surface

Xueyuan Wu and Asok K. Ray*

*P. O. Box 19059, Physics Department, The University of Texas at Arlington
Arlington, Texas 76019*

The electronic and geometric properties of bulk fcc δ-Plutonium and the quantum size effects in the surface energies and the work functions of the (001) ultra thin films (UTF) up to 7 layers have been investigated with periodic density functional theory calculations within the full-potential linearized augmented-plane wave (FP-LAPW) approach as implemented in the WIEN2k package. The effects of several approximations have been examined: (1) non-spin polarization (NSP) vs. spin polarization (SP); (2) scalar-relativity (no spin-orbit coupling (NSO)) vs. full-relativity (i.e., with spin-orbit (SO) coupling included). Our calculations show that both spin-polarization and spin-orbit coupling play important roles in determining the equilibrium atomic volume and bulk modulus for δ-Plutonium. Our calculated equilibrium atomic volume of 178.3 a.u.$^3$ and bulk modulus of 24.9 GPa at the fully relativistic level of theory, *i.e.* spin-polarization and spin-orbit coupling included, are in good agreement with the experimental values of 168.2 a.u.$^3$ and 25 GPa (593°K), respectively. In particular, the energy difference brought by spin-orbit coupling, about 7-8 eV, is dominant, but the energy difference brought by spin-polarization, from a few tenths to 2 eV, has a stronger dependence on the atomic volume. Features of the density of states show that 5f electrons are more itinerant when the volume of δ-plutonium is compressed and they are more localized when the volume is expanded, which provides evidence to explain the origin of the volume expansion between the α- and δ-phases. The calculated equilibrium lattice constants at different levels of approximation are used in the surface properties calculations for the thin films. The surface energy is found to be rapidly converged at all four levels approximation, NSP-NSO, NSP-SO, SP-NSO, and SP-SO. The semi-infinite surface energy is predicted to be 0.692eV at the fully-relativistic level with spin-polarization and spin-orbit coupling. Quantum size effect for the work function is not found to be pronounced for the (001) surface at the LAPW level of theory.

akr@uta.edu

## 1. Introduction

Recent years have seen increased interests in the studies of the strongly correlated and heavy fermion systems, including the actinides.[1-6] As is known, the actinides are characterized by a gradual filling of the 5f-electron shell with the degree of localization increasing with the atomic number Z along the last series of the periodic table. The open shell of the 5f electrons determines the magnetic and solid-state properties of the actinide elements and their compounds and understanding the quantum mechanics of the 5f electrons is the defining issue in the physics and chemistry of the actinide elements. These elements are also characterized by the increasing prominence of relativistic effects and their studies can, in fact, help us to understand the role of relativity throughout the periodic table. Narrower $5f$ bands near the Fermi level, compared to $4d$ and $5d$ bands in transition elements, is believed to be responsible for the exotic nature of actinides at ambient condition.[3] The 5f orbitals have properties intermediate between those of localized 4f and delocalized 3d orbitals and as such, the actinides constitute the "missing link" between the d transition elements and the lanthanides.[1] Thus a proper and accurate understanding of the actinides will help us understand the behavior of the lanthanides and transition metals as well.

Among the actinides, the unique electronic properties of plutonium (Pu), which does not occur naturally, but is man-made for nuclear power and other related purposes, have generated considerable interest in recent years, from both scientific and technological points of view. The special location of Pu in the actinide series, on the border between light actinides characterized by metallic-like 5f electron behavior and the heavy actinides characterized by localized 5f electron behavior, is believed to be responsible for its complex phase diagrams containing at least six stable allotropes as a function of temperature.[7] In particular, Pu undergoes a twenty-five percent volume expansion from its ambient α-phase monoclinic structure with sixteen atoms per unit cell and stable below 400 K to the δ-phase with a face-centered-cubic (fcc) structure and stable above 600 K.[8] However, the high-temperature fcc δ-phase of Pu can be stabilized at room temperature through the addition of very small amounts of certain impurities. For example, $Pu_{1-x}Ga_x$ has the fcc structure and physical properties of δ-Pu for $0.020 \leq x \leq 0.085$.[9] In fact, δ-Pu exhibits properties that are intermediate between the light- and heavy-actinides. The α-phase Pu with atomic volume aligning with the other light actinides (Th to Np) is relatively accurately described by standard density functional theory (DFT) [10] electronic structure calculations.[11] However, the δ-phase Pu, though technologically more important, is less understood. DFT in its local density or generalized gradient approximations (LDA or GGA) [12] generally fails to predict the correct atomic volume and bulk modulus and experimental photoemission spectra for δ-Pu unless some form of magnetic ordering, *e.g.* collinear ferromagnetism is considered.[13-14] In fact, conventional DFT typically underestimates the equilibrium atomic volume of δ-Pu by 20-30%, and overestimates bulk modulus by about 300%. [15-16] The validity of DFT in dealing with the localized highly correlated 5f electrons has been a matter of significant controversy, since the 5f electrons in the fcc δ-phase Pu are partially localized[17] as indicated by its atomic volume which is approximately halfway between α-Pu and heavy actinide americium. To remedy the problems, corrections have been introduced based on different approaches such as dynamical mean-field theory, [17] muffin tin orbital calculations, [18] spin and orbital polarizations. [13,19-22] Different approaches have yielded different degrees of success in dealing with δ - Pu. For example, the LDA+U method, [23-25] where U is the adjustable Hubbard parameter, to describe the electron correlation within the dynamical mean field theory (DMFT) has been used to study δ - Pu. The experimental equilibrium δ-Pu volume was reproduced, with U equal to 4 eV. Penicaud[18] performed total energy calculations in the local density approximation using fully relativistic muffin-tin orbital band structure method. For δ-Pu, the $5f_{5/2}$ electrons are uncoupled from the s, p and d electrons to reproduce experimental value of equilibrium atomic volume. Also an adjustable parameter was introduced to get better theoretical representation of δ-Pu. Using 'mixed-level' model, where the energies were calculated at both localized and delocalized 5f configurations, Eriksson *et al.*[26] reproduced reasonable equilibrium volumes of U, Pu and Am. Theories beyond LDA, such as, the self-interaction-corrected (SIC) LDA studied by Petit *et al.*[27] predicted a 30% too large equilibrium volume.

Apart from equilibrium volume and bulk modulus, the existence of magnetic moments in bulk δ-Pu is also a subject of great controversy and significant discrepancies exist between various experimental and theoretical results. To this end, we comment on a *few* representative works in the literature, partly to mention explicitly some of the controversies. Susceptibility and resistivity data for δ-Pu were published by Meot-Reymond and Fournier, [28] indicating the existence of small magnetic moments screened at low temperatures. This screening was attributed to the Kondo effect. Recent experiments by Curro and Morales[9] of 1.7 percent Ga-doped δ-Pu conducted at temperatures lower than the proposed Kondo Temperature of 200-300 K showed little evidence for local magnetic moments at the Pu sites. Though there is no direct evidence for magnetic moment, spin-polarized DFT, specifically the generalized-gradient-approximation (GGA) to DFT, has been used by theoreticians, in particular, to predict the magnetic ordering and the ground state properties of δ-Pu. This is partly due to the fact that spin-polarized DFT calculations do predict better agreement with photoemission data. Niklasson *et al.*[29] have presented a first- principles disordered local moment (DLM) picture within the local-spin-density and coherent potential approximations (LSDA+CPA) to model some of the main characteristics of the energetics of the actinides, including δ-Pu. The authors also described the failures of the local density approximation (LDA) to describe 5f localization in the heavy actinides, including elemental Pu and the DLM density of states was found to compare well with photoemission on δ-Pu, in contrast to that obtained from LDA or the magnetically ordered AFM configuration. On the other hand, Wang and Sun[30], using the full-potential linearized augmented-plane-wave (FP-LAPW) method within the spin-polarized generalized gradient approximation (SP-GGA) to density functional theory, without spin-orbit coupling, found that that the anti-ferromagnetic state lattice constant and bulk modulus agreed better with experimental values than the nonmagnetic values of δ-Pu. Using the fully relativistic linear combinations of Gaussian-type orbitals-fitting function (LCGTO-FF) method as embodied in the program GTOFF[31], (the current version allowing for simultaneous inclusion of spin-polarization and spin-orbit coupling in the screened-nuclear-spin-orbit SNSO approximation) Boettger[21] found that, at zero pressure, the AFM (001) state was bound relative to the non-magnetic state by about 40 mRy per atom. The lattice constant for the AFM (001) state also agreed better with the experimental lattice constant as compared to the nonmagnetic lattice constant. However, the predicted bulk modulus was significantly larger than the experimental value. Söderlind *et al.*[13, 20, 32], employing all electron, full-potential-linear-muffin-tin-orbitals (FLMTO) method, predicted a mechanical instability of anti-ferromagnetic δ-Pu, and proposed that δ-Pu is a 'disordered magnet'. In a more recent study on 5f localization, Söderlind *et al.* showed that 5f-band fractional occupation at 3.7 (68% atoms with itinerant 5f electrons) predicts well the atomic volume and bulk modulus without referring to the magnetic ordering. Wills *et al.*[33] have claimed that there is, in fact, no evidence of magnetic moments in the bulk δ-phase, either ordered or disordered. Although the magnetic order of δ-Pu remains controversial, the spin-polarization that gives rise to the formation of a spin moment is a fundamental issue and cannot be ignored. Otherwise, DFT fails to predict any known properties of δ-Pu. One of the objectives of this work is thus to investigate the precise and comparative roles of spin-polarization and spin-orbit coupling in the bulk properties of δ-Pu.

Apart from the volume expansion induced by phase transition of Pu, the surface corrosion of Pu in the presence of environmental gases is another challenging problem, crucial for issues of long-term storage and disposal. Thus, the second phase of this work is to study the surface electronic properties of δ-Pu, specifically the (001) surface providing the basis for further understanding of the chemical reactivity of Pu and of surface corrosion. The unusual aspects of the bonding in bulk Pu are apt to be enhanced at a surface or in a thin layer of Pu adsorbed on a substrate, as result of the reduced atomic coordination of a surface atom and the narrow bandwidth of surface states. Thus, Pu surfaces and thin films may also provide valuable information about the bonding in Pu. Grazing-incidence photoemission studies, combined with the calculations of Eriksson *et al.*[34], suggest the existence of a small-moment δ-like surface on α-Pu. Using GTOFF, Ray and Boettger[35] have also indicated the possibility of such a surface for a Pu monolayer. Recently, high-purity ultra-thin layers of Pu deposited on Mg were studied by X-ray photoelectron (XPS) and high-resolution valence band (UPS) spectroscopy by Gouder *et al.*[36] They found that the degree of delocalization of the 5f states depends in a very dramatic way on the layer thickness and that the itinerant character of the 5f states is gradually lost with reduced thickness, suggesting that the thinner films are δ-like. Finally, it may be possible to study 5f localization in Pu layers through adsorption on a series of carefully selected substrates; in which case, the adsorbed layers are more likely to be δ-like than α– like. . We also wish to mention that *no detailed information exists in the literature about the magnetic state of the surface of δ -Pu* and our present study including spin polarization *and* spin-orbit coupling on Pu surfaces is a *first step* towards an understanding of Pu surfaces *and* the influence of magnetism on such surfaces. We also note that, as the films get thicker, the complexity of magnetic ordering, if existent, increases and such calculations are quite challenging computationally. Nevertheless, to study the effects of spin polarization on the surface electronic structure, our studies have been performed at both the spin-polarized and at the non-spin-polarized levels.

For studies like these, it is common practice to model the surface of a semi-infinite solid with an ultra-thin film (UTF), which is thin enough to be treated with high-precision density functional calculations, but is thick enough to model the intended surface realistically. Determination of an appropriate UTF thickness is complicated by the existence of possible quantum oscillations in UTF properties as a function of thickness; the so-called quantum size effect (QSE). These oscillations were first predicted by calculations on jellium films [37,38] and were subsequently confirmed by band structure calculations on free-standing UTFs composed of discrete atoms.[39-42] The adequacy of the UTF approximation obviously depends on the size of any QSE in the relevant properties of the model film. Thus, it is important to determine the magnitude of the QSE in a given UTF prior to using that UTF as a model for the surface. This is particularly important for Pu films, since the strength of the QSE is expected to increase with the number of valence electrons.[37] Using GTOFF, [31] a recent fully relativistic first principles study of the ultra-thin (111) films of fcc δ-Pu up to five layers thick showed that while the surface energies converge within the first three layers, the work function exhibits a strong QSE.[43] This work is, in some sense, an extension of the previous work, using *albeit* a different computational formalism, for the (001) films of fcc δ-Pu. The formalism used here an all-electron full-potential calculation with mixed basis set of linearized augmented-plane wave (LAPW) and augmented-plane-wave plus local orbitals (APW + lo), with and without spin-orbit coupling (SO), [44-46] to study both the bulk properties of δ-Pu and the thickness dependence of the surface properties for the (001) surface.

## 2. Computational Method

The computations have been carried out using the full-potential all-electron method with mixed basis APW+lo / LAPW method implemented in the WIEN2k suite of programs.[46] A gradient corrected Perdew- Berke - Ernzerhof (PBE) [47] functional is used to describe the exchange and correlation effects. In the WIEN2k code, the alternative basis set APW+lo is used inside the atomic spheres for the chemically important orbitals that are difficult to converge, whereas LAPW is used for others. The local orbitals scheme leads to significantly smaller basis sets and the corresponding reductions in computing time, given that the overall scaling of LAPW and APW + lo is given by $N^3$, where N is the number of atoms. Also, results obtained with the APW + lo basis set converge much faster and often more systematically towards the final value.[48] As far as relativistic effects are concerned, core states are treated fully relativistically in WIEN2k and for valence states, two levels of treatments are implemented: (1) a scalar relativistic scheme that describes the main contraction or expansion of various orbitals due to the mass-velocity correction and the Darwin s-shift[49] and (2) a fully relativistic scheme with spin-orbit coupling included in a second-variational treatment using the scalar-relativistic eigenfunctions as basis.[50,51] For the bulk calculations, an fcc unit cell with one atom is used. To calculate the total energy at $0^o$ K, a constant muffin-tin radius ($R_{mt}$) of 2.70 a.u. is used for all atomic volumes. The plane-wave cut-off $K_{cut}$ is determined by $R_{mt} K_{cut}$ = 9.0. The Brillouin zone is sampled on a uniform mesh with 104 irreducible K-points. The (001) surface of fcc $\delta$-Pu is modeled by periodically repeated slabs of N Pu layers (with one atom per layer and N=1-7) separated by a 60 Bohr a.u. vacuum gap. Twenty-one irreducible K points have been used for reciprocal-space integrations. For each calculation, the energy convergence criterion is set to be 0.006 mRy.

## 3. Results and Discussions

### 3.1 The bulk properties of fcc $\delta$-Pu

We have calculated the total energies for a set of different atomic volumes and the results are presented in Figure 1 at both NSP and SP levels, with and without spin-orbit coupling. The Murnaghan equation of state[52]

$$E = B \times V / \beta \times (1/(\beta-1) \times (V_0/V)^\beta + 1) \tag{1}$$

is used to fit the total energy curve and obtain the equilibrium atomic volume and the bulk modulus. Results are listed in Table 1 and compared with some of the available theoretical and experimental results.[13,14,20,21,23,30,53,54] Our results show that the equilibrium atomic volume of δ-Pu is underestimated by 21% and 27%, with and without SO coupling, respectively, compared to the experimental value without spin-polarization and the bulk modulus is considerably overestimated by the non-spin-polarized calculations. Inclusion of spin-polarization significantly improves the values of both the atomic volume and the bulk modulus compared to the experimental values. For instance, our calculated bulk moduli at the level of SP-NSO and SP-SO are 32.5 GPa and 24.9 GPa, respectively, which are very close to the experimental value of 25 GPa at 593.1° K.[54] The 0°K experimental bulk modulus deduced from linear fits to data listed in Ref. 54 is 35.1 GPa.[21] Our equilibrium atomic volume of 178.3 (a.u.[3]) and bulk modulus of 24.9 GPa at the SP-SO level are in good agreement with the LCGTO-FF-SO-FM values of 176.2 (a.u.[3]) and a bulk modulus of 31.3 GPa. Also, the magnetic moment obtained at the SP-SO level is 5.05 $\mu_B$, to be compared with the LCGTO-FF-SO-FM value of 5.1 $\mu_B$.[21] From the results we can see that both spin-polarization and spin-orbit coupling play important roles in predicting the bulk properties. For example, SP brings the bulk modulus from 214.2 GPa down to 32.5 GPa without SO included, while the SO brings the bulk modulus from 214.2 GPa down to 101.9 GPa without considering spin-polarization. The small effective moment given by recent magnetic-susceptibility data for alloy-stabilized δ-Pu and *some* other theoretical investigations as mentioned above suggesting that a large anti-parallel orbital moment almost cancel the spin moment, also support the idea that spin-polarization needs to be taken into account in dealing with δ-Pu, a highly correlated system.

To further assess the effects brought by SP and SO, we plot in Figure 2 spin polarization energy $E_{SP}$ and spin-orbit coupling energy $E_{SO}$, respectively as a function of atomic volume. The spin-polarization energy $E_{SP}$ is defined by

$$E_{SP} = E_{tot}(NSP) - E_{tot}(SP) \tag{2}$$

and

$$E_{SO} = E_{tot}(NSO) - E_{tot}(SO) \tag{3}$$

where $E_{tot}$ (NSP) and $E_{tot}$ (SP) are the total energy at the NSP and SP levels, respectively; $E_{tot}$ (NSO) and $E_{tot}$ (SO) are the total energy at the NSO and SO levels, respectively. It can be seen from Figure 2 that $E_{SO}$ (about 7-8 eV) is much larger than $E_{SP}$ (about a few tenths to about 2.0 eV). We also note that the LAPW method uses the well-known variational collapse problem because of the way the basis set is formed and the second-variational treatment is expected to slightly underestimate the effects of spin-orbit coupling. However, linear fits show that $E_{SP}$ has a larger slope (0.022 eV/Bohr$^3$ for NSO and 0.011 eV/Bohr$^3$ for SO levels, respectively) compared to $E_{SO}$ (0.009eV/Bohr$^3$ for NSP and -0.001 eV/Bohr$^3$ for SP level, respectively). Hence, spin-polarization does play a dominant role in determining the equilibrium atomic volume.

To understand how the atomic volume affects the nature of the 5f electrons in Pu metal, we plot in Figure 3 the density of states (DOS) of 5f electrons for three different atomic volumes: the calculated equilibrium volume and the largest and smallest volumes considered in our SP-SO calculations. We first note the relatively narrow 5f electron bands located around the Fermi level set at zero in the DOS. However, it is also obvious that some of the 5f electrons are localized and some are itinerant, agreeing with some studies and in disagreement with others.[13,20,32,33] It's interesting to see from the position of the Fermi level, that the 5f electron bandwidth beyond the Fermi level increases as the atomic volume is compressed. This suggests that the degree of localization of the 5f electrons related to the bonding of Pu metal is closely related with the volume expansion, namely, the 5f electrons are more itinerant if the volume is compressed and more localized if the volume is expanded. This provides evidence explaining the origin of the volume expansion when the Pu metal undergoes transition from α-phase to δ-phase. This also supports the dynamical mean-field theory study of Savrasov et al.[17] which suggests that the α-phase and the δ-phase are on opposite sides of the interaction-driven delocalization-localization transition.

### 3.2 Properties of δ-Pu (001) thin films

We now consider ultra-thin-films (UTF) of the (001) surface of δ - Pu. For each film up to seven layers thick, we have calculated cohesive energies per atom, surface energies, and work functions at *both non-spin-polarized and polarized levels, with and* *without spin orbit coupling* (tables 2-3). Because of severe demands on computational resources and internal consistency, the calculated equilibrium lattice constants obtained at different levels of approximation for bulk Pu have also been chosen in the slab computations at the corresponding level of approximation. Spin polarization energy per atom, spin orbit coupling energy per atom and total magnetic moment per atom have also been calculated and the results are summarized in table 4.

The cohesive energies $E_{coh}$ per layer for the N-layer slabs have been calculated with respect to N monolayers and are found to increase monotonously with the thickness (see Figure 4) at all four levels of calculations. These features are in general agreement with the results of Ray and Boettger [43] for the (111) surface of δ-Pu using the linear combinations of Gaussian type orbitals –fitting function (LCGTO-FF) method. For the thickest layer considered in the LCGTO-FF study, namely a 5-layer (111) film had a cohesive energy of 1.58eV at the NSP-SO level, to be compared with our corresponding result of 1.87eV for the 5-layer (001) film. The differences can be attributed to the different surfaces and the different methodology used. The LAPW method, a computationally less intensive method compared to the LCGTO-FF method, uses a diffuse orbital basis set and the basis functions can have discontinuities in their second derivatives at the muffin-tin sphere boundaries. On the other hand, the LCGTO-FF method uses a local basis set without the existence of discontinuities. In this study, we also find that the rate of increase of cohesive energy drops significantly as the number of layers increase and it is expected that saturation can be achieved after some more layers. However, since the experimental value for the semi-infinite surface cohesive energy is not known to the best of our knowledge, we are unable to predict how many layers will be needed to achieve the bulk surface energy. We also note that spin-polarization lowers the cohesive energy by about 45-51% at the scalar relativistic level and by about 30-35% at the SO level. The SO coupling increases the cohesive energy of the spin-polarized N-layers by about 8-12% but reduces the cohesive energy of the non-spin-polarized N-layers by about 14-16%. The incremental energies $E_{inc}$ of N-layers with respect to (N-1)-layers plus a single monolayer are also calculated and plotted in Figure 5. $E_{inc}$ is found to be quickly saturated. The non-spin-polarized calculations do show pronounced oscillations in $E_{inc}$ for $N \leq 3$. At the spin-polarized levels, with and without SO, $E_{inc}$ is about 0.1eV from the di-layer to the tri-layer; however, these oscillations are basically non-existent beyond these in the spin-polarized calculations. The LAPW incremental energies are rather close to the GTOFF values. For the 5-layer (111) film, the GTOFF value is 1.93eV, as compared with a value of 2.24eV at the NSP-SO level for the 5-layer (001) film.

The surface energy for an N-layer film may be estimated from: [55]

$$E_s = (1/2) [E_{tot}(N) - N E_B] \tag{4}$$

where $E_{tot}(N)$ is the total energy of the N-layer slab and $E_B$ is the energy of the infinite crystal. If N is sufficiently large and $E_{tot}(N)$ and $E_B$ are known to infinite precision, Eq. (4) is exact. If, however, the bulk and the film calculations are not entirely consistent with each other, $E_s$ will diverge linearly with increasing N.[56] Stable and internally consistent estimates of $E_s$ and $E_B$ can, however, be extracted from a series of values of $E_{tot}(N)$ versus n via a linear least-squares fit to: [56]

$$E(N) = E_B N + 2E_s \tag{5}.$$

To obtain an optimal result, the fit to Eq. (5) should only be applied to films which include, at least, one bulk-like layer, *i.e.* N>2. We have independently applied this procedure to the films at all four levels of theory. At the NSP-NSO level, we obtain $E_B$ = -59384.02969 Ry and $E_s$ = 1.01 eV, and $E_B$ = -59384.59085 Ry. and $E_s$ = 0.94 eV at the NSP-SO level. At the SP-NSO level, $E_B$ = -59384.08827 Ry and $E_s$ = 0.73 eV, and at the SP-SO level, the corresponding values are –59384.61766 Ry and $E_s$ = 0.69 eV. Thus, the semi-infinite surface energy decreases by close to thirty-one percent from the non-spin-polarized-scalar-relativistic case to the spin-polarized-fully relativistic case. The surface energy for each layer has been computed using the calculated N-layer total energy and the appropriately fitted bulk energy. Results are listed in table 3 and plotted in figure 6. The surface energies obtained with non-spin-polarized calculations with or without SO coupling oscillate up to 3 layers and tend to converge for larger N. The surface energies obtained with spin-polarized calculations with or without SO show a small drop from monolayer to di-layer and are pretty well converged for thicknesses beyond 2 layers. *From these results, we again infer, similar to our conclusions for the (111) surface using GTOFF, that a 3-layer film might indeed be sufficient if the primary quantity of interest is the chemisorption energy.*[43] It is worth noting though that because of severe demands on computational resources, the GTOFF results up to 5 layers were performed without spin polarization and that we were able to treat only the monolayer and the di-layer at all four levels of theory.

We have also investigated the dependence of the work function on the layer thickness. The work function W is calculated from

$$W = V_0 - E_F \tag{6},$$

and $V_0$ is the Coulomb potential energy at half Z where Z is the height of the slab including the vacuum layer. $E_F$ is the Fermi energy. The results are listed in Table 3 and plotted in Figure 7. Some oscillations are observed at all levels but they are not pronounced at the SP-SO level of theory. Unlike the (111) film, for which a clear even-odd oscillatory pattern is shown in the work function, [43] this kind of QSE is not evident for the (001) slabs in the FP-LAPW calculations. Hence, quantum size effect does not appear to be highly significant and we can use a relatively thin film to model the (001) surface for our future investigations on adsorption studies on this surface. The average work functions we obtained here are about 3.4 eV for NSP and 3.0 eV for SP calculations with or without SO. These values are in fair agreement with the value of 3.68eV obtained by Hao *et al.*[34]

We also calculated the spin-polarization energies, spin-orbit coupling energies and magnetic moments per atom for each N-layer slab. The results are listed in table 4, which shows a quick convergence as well. The fairly well converged values of spin-polarization energy of 0.9 eV and 0.4 eV at scalar and fully relativistic levels, respectively, are comparable with Boettger's results of 0.73 eV and 0.37 eV for ferromagnetic bulk δ-Pu at corresponding levels.[21] The scalar- and fully- relativistic calculations yield spin-orbit energies of 7.65eV and 7.21 eV, respectively and the magnetic moments per atom are 5.9 and 5.3 $\mu_B$, respectively with and without SO.

In Figure 8, we plot density of states of Pu 5f electrons for bulk Pu and (001) N-layer slabs, where N=1, 3, 7. For the N-layers, except for the second panel from above where DOS of the center layer Pu 5f states are plotted, others are for the top layer Pu 5f electrons. Several features are observed: a) for the monolayer, one can see the splitting brought by SO, which also increases the 5f bandwidth; b) as the thickness of the slab increases from N=1 to N=3, more 5f states are beyond the Fermi level, which indicates an increasing degree of delocalization with respect to the thickness. However, from N=3 to N=7, the 5f bandwidth beyond the Fermi level slightly reduced. This indicates the thickness dependence of the degree of 5f delocalization is varying and this is consistent with X-ray photoelectron and high-resolution valence band spectroscopic data;[36] c) DOS for the monolayer shows that the 5f states are mostly localized. For the 7-layer slab, the DOS is also plotted for Pu 5f electrons in the center layer, which shows features very close to that of the bulk DOS.

## 4. Conclusions

Full-potential all-electron density functional calculations with mixed basis APW+lo / LAPW implemented in the WIEN2k packet have been carried out to investigate the electronic structures of both bulk δ-Pu and ultra thin (001) films at both scalar- and fully- relativistic levels, with and without spin-polarization. We find that DFT is able to fairly well reproduce the known properties including the equilibrium atomic volume and bulk modulus of δ-Pu if spin-polarization and spin-orbit coupling are taken into account. Although the magnetic state of δ-Pu remains unclear, we find that SP plays an important role in determining the bulk properties and the spin-polarization energy has a strong dependence on the atomic volumes. Features of the density of states show that 5f electrons are more itinerant when the volume of δ-Plutonium is compressed and they are more localized when the volume is expanded, which provides evidence to explain the origin of the volume expansion between the α- and δ-phases.

The thickness dependence of work functions, surface energies, cohesive energies, incremental energies, magnetic moments, spin-polarization and spin-orbit coupling energies per atom have been also studied at the four levels of approximations, namely NSP-NSO, NSP-SO, SP-NSO, and SP-SO. We find that an ultra-thin film (3 layers thick) can be used to model the (001) surface of δ-Pu to a good approximation. The DOS feature of the 5f states on top layer of N-layer films shows a varying thickness dependence of the degree of 5f delocalization. The itinerant character of the 5f electrons is lost in the monolayer. These results are, in general, consistent with experimental observations.

## Acknowledgements

This work is supported by the Chemical Sciences, Geosciences and Biosciences Division, Office of Basic Energy Sciences, Office of Science, U. S. Department of Energy (Grant No. DE-FG02-03ER15409) and the Welch Foundation, Houston, Texas (Grant No. Y-1525).

Table 1. Calculated equilibrium atomic volume $V_0$ (a.u.$^3$) and bulk modulus B (GPa)

| Method | $V_0$ (a.u.$^3$) | B(GPa) |
|---|---|---|
| PBE-NSP-NSO [a] | 122.8 | 214.2 |
| PBE-NSP-SO [a] | 132.9 | 101.9 |
| PBE-SP-NSO [a] | 188.4 | 32.5 |
| PBE-SP-SO [a] | 178.3 | 24.9 |
| LCGTO-FF-NSP-SO [b] | 141.4 | 97 |
| FLAPW-GGA-SO [b] | 133.4 | 121 |
| LCGTO-FF-SO-FM [c] | 176.2 | 31.3 |
| LCGTO-FF-SO-AFM (100) [c] | 162.9 | 53.5 |
| LDA+ U [d] | 170.3 | 61 |
| FLAPW-GGA-CMF-AFM [e] | 168.7 | 43 |
| FLMTO [f] | 182.3 | 21 |
| Exp | 168.2[g] | 25[h] |

(a) This work

(b) Ref. 14

(c) Ref. 21, FM (001) ferromagnetic; AFM (100) anti-ferromagnetic.

(d) Ref. 23

(e) Ref. 30, CMF (classical mean field statistics).

(f) Ref. 13,20

(g) Ref. 53

(h) Ref. 54, modulus at 593.1$^o$K.

Table 2. Cohesive energies $E_{coh}$ per atom with respect to the monolayer and incremental energies $E_{inc}$ for δ-Pu (001) N-layers (N=1-7). All energies are in eV.

| N | $E_{coh}$ (eV) | | | | $E_{inc}$ (eV) | | |
|---|---------|--------|--------|-------|---------|--------|-------|
|   | NSP-NSO | NSP-SO | SP-NSO | SP-SO | NSP-NSO | NSP-SO | SP-NSO | SP-SO |
| 2 | 1.59 | 1.34 | 0.77 | 0.86 | 3.18 | 2.67 | 1.55 | 1.73 |
| 3 | 1.91 | 1.61 | 1.01 | 1.12 | 2.55 | 2.16 | 1.48 | 1.63 |
| 4 | 2.09 | 1.77 | 1.13 | 1.24 | 2.66 | 2.26 | 1.49 | 1.62 |
| 5 | 2.19 | 1.87 | 1.2 | 1.31 | 2.56 | 2.24 | 1.5 | 1.58 |
| 6 | 2.25 | 1.93 | 1.25 | 1.36 | 2.59 | 2.23 | 1.47 | 1.59 |
| 7 | 2.3 | 1.97 | 1.29 | 1.39 | 2.58 | 2.23 | 1.51 | 1.57 |

Table 3. Surface energies and work functions for δ-Pu (001) N-layers (N=1-7). All energies are in eV.

| | $E_s$(eV) | | | | W (eV) | | | |
|---|---|---|---|---|---|---|---|---|
| N | NSP-NSO | NSP-SO | SP-NSO | SP-SO | NSP-NSO | NSP-SO | SP-NSO | SP-SO |
| 1 | 1.30 | 1.12 | 0.75 | 0.79 | 3.38 | 3.37 | 3.02 | 3.1 |
| 2 | 1.00 | 0.91 | 0.72 | 0.72 | 3.55 | 3.52 | 2.99 | 3.08 |
| 3 | 1.02 | 0.95 | 0.73 | 0.70 | 3.41 | 3.48 | 2.88 | 3.07 |
| 4 | 0.99 | 0.94 | 0.73 | 0.69 | 3.39 | 3.44 | 3.04 | 3.04 |
| 5 | 1.00 | 0.94 | 0.72 | 0.69 | 3.41 | 3.43 | 2.91 | 3.01 |
| 6 | 1.01 | 0.95 | 0.73 | 0.69 | 3.41 | 3.35 | 2.98 | 3.04 |
| 7 | 1.04 | 0.92 | 0.72 | 0.70 | 3.47 | 3.48 | 2.95 | 3.02 |

Table 4. Spin-polarization energies per atom $E_{SP}$ and spin-orbit coupling energies per atom $E_{SO}$. All energies are in eV. Also listed are the magnetic moments per atom (in Bohr magnetons) MM for δ-Pu (001) N-layers (N=1-7).

| | $E_{SP}$ (eV) | | $E_{SO}$ (eV) | | MM ($\mu_B$) | |
| N | NSO | SO | NSP | SP | NSO | SO |
| --- | --- | --- | --- | --- | --- | --- |
| 1 | 1.9 | 1 | 7.99 | 7.11 | 7.1 | 6.6 |
| 2 | 1.1 | 0.5 | 7.73 | 7.2 | 6.2 | 5.6 |
| 3 | 1 | 0.5 | 7.69 | 7.22 | 6.1 | 5.4 |
| 4 | 0.9 | 0.5 | 7.66 | 7.22 | 5.9 | 5.5 |
| 5 | 0.9 | 0.5 | 7.66 | 7.22 | 5.9 | 5.3 |
| 6 | 0.9 | 0.4 | 7.66 | 7.22 | 5.9 | 5.2 |
| 7 | 0.9 | 0.4 | 7.65 | 7.21 | 5.9 | 5.3 |

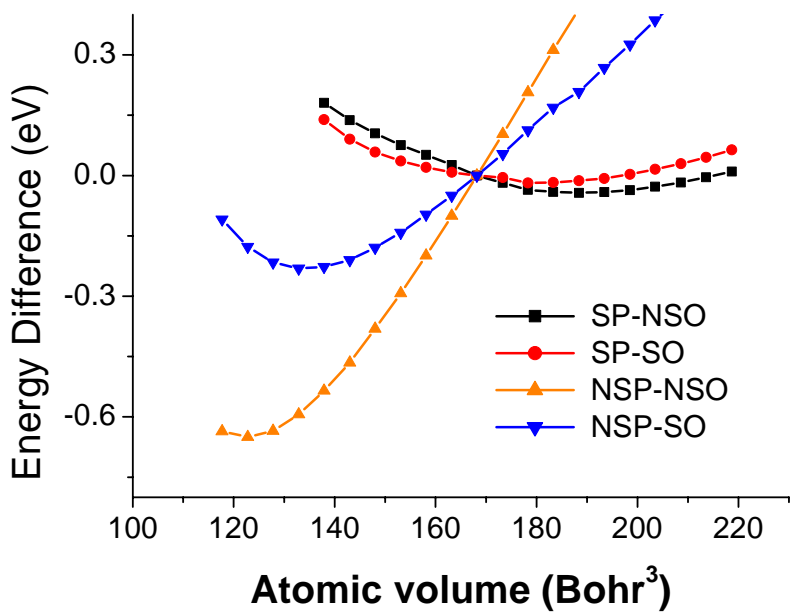

Fig.1 The energy difference $\Delta E = E_V - E_{V_0}$ vs. atomic volume V, where $E_V$ is the $0^\circ K$ total energy at volume V, $V_0$ is the experimental atomic volume.

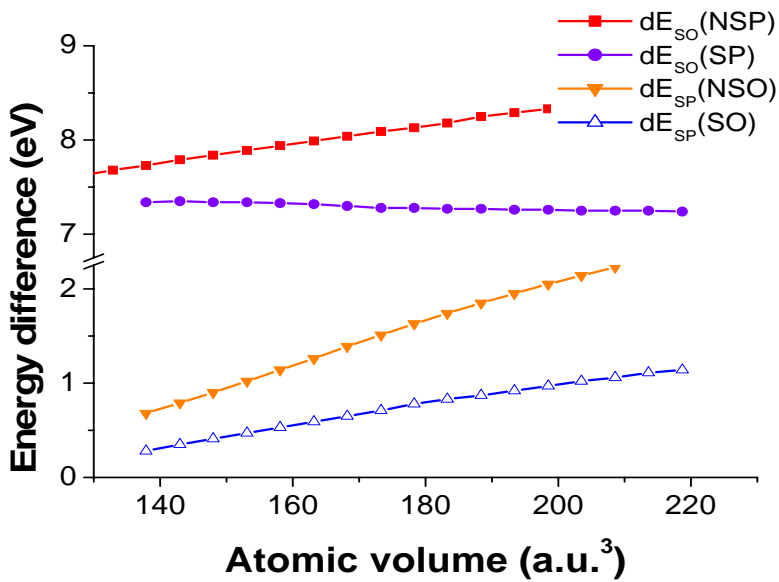

Fig.2. Spin polarization energy $E_{SP}$ and spin orbit coupling energy $E_{SO}$ versus atomic volume.

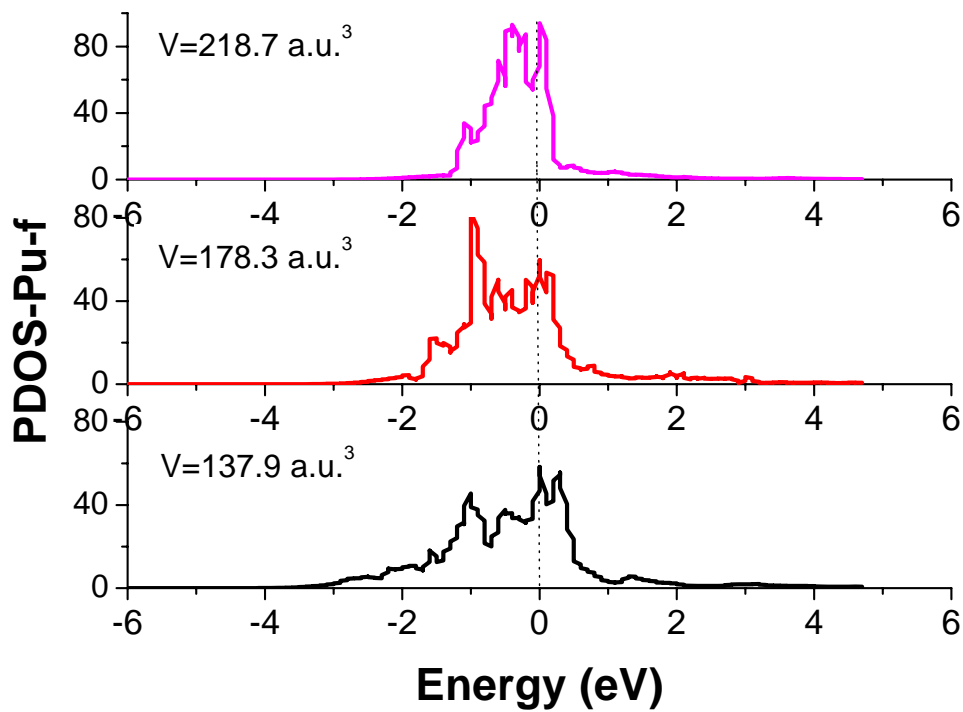

Figure 3. Projected density of states on Pu 5f electrons at different atomic volumes obtained by SP-SO calculations. Fermi energy is set at zero.

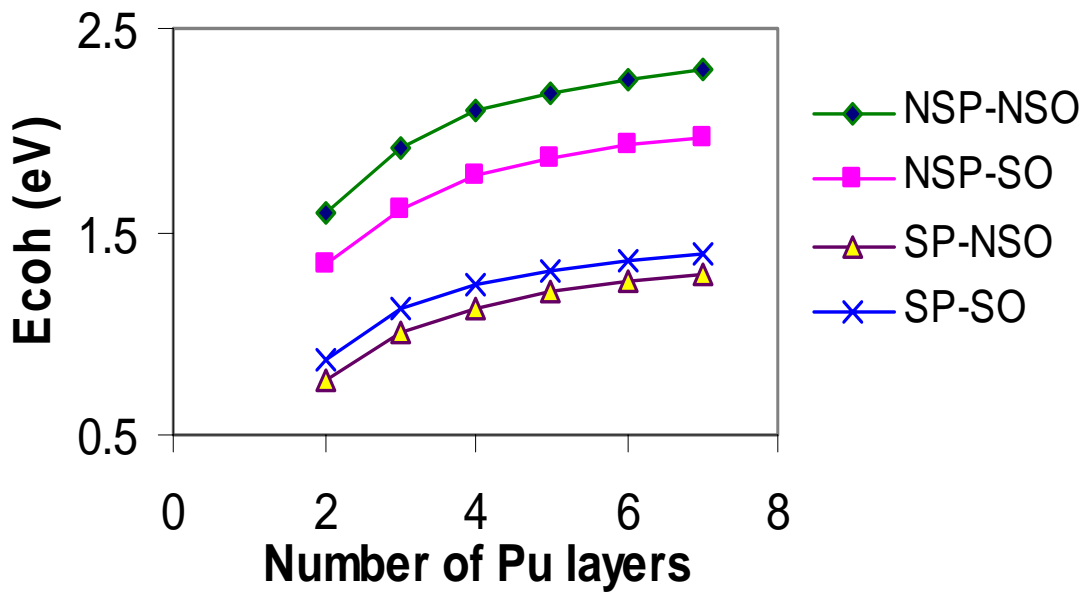

Figure 4. Cohesive energy per atom versus number of Pu layers.

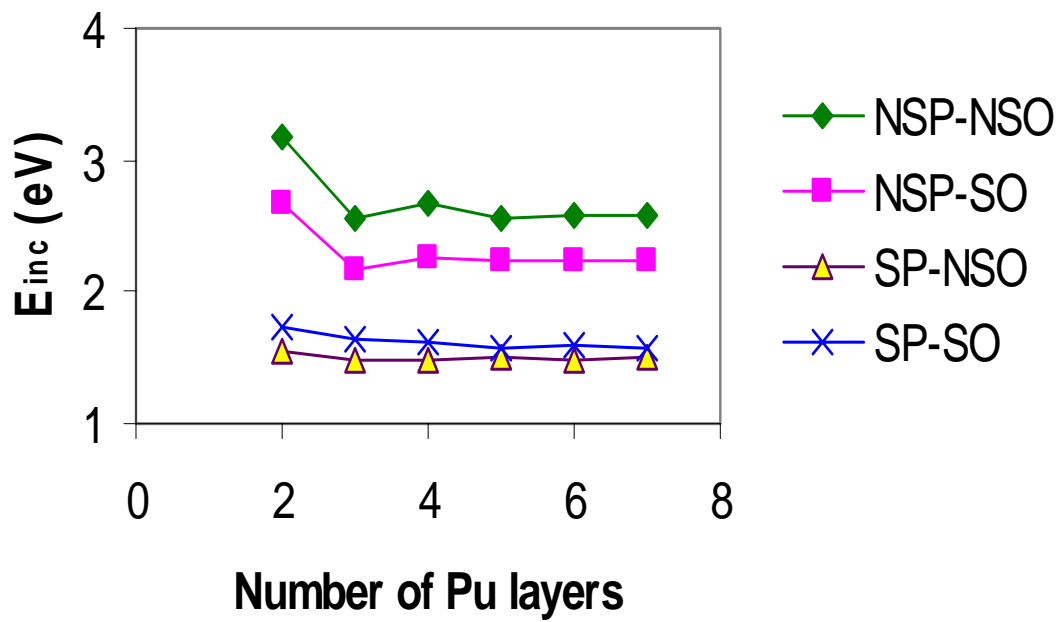

Figure 5. Incremental energy versus number of Pu layers.

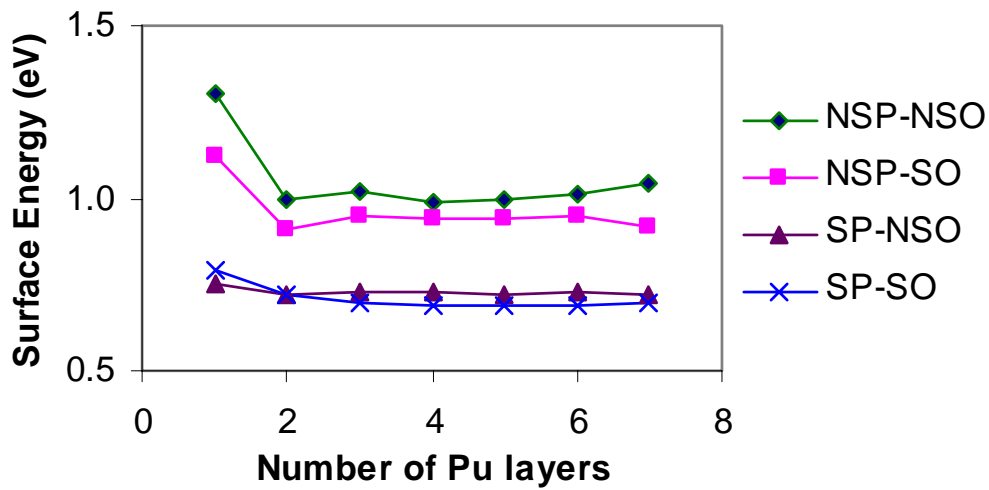

Figure 6. Surface energy versus number of Pu layers.

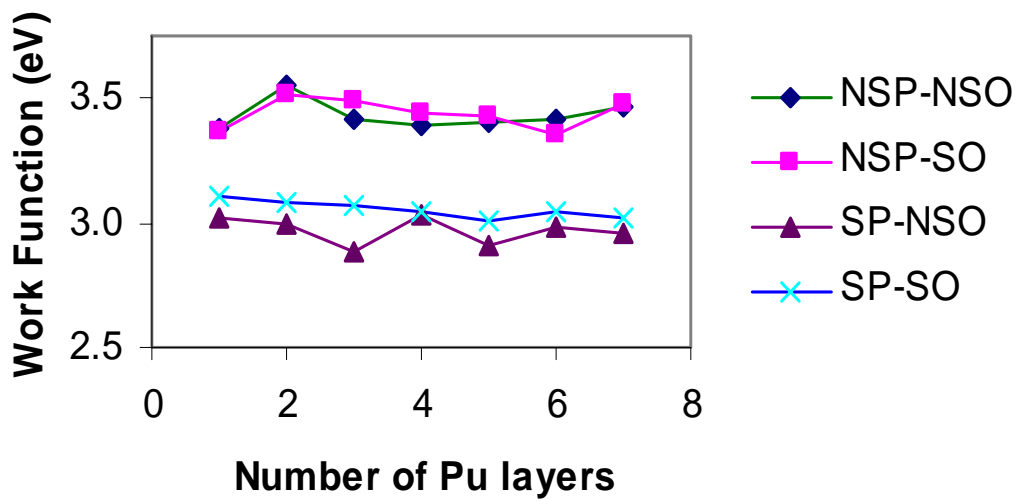

Figure 7. Work function versus number of Pu layers.

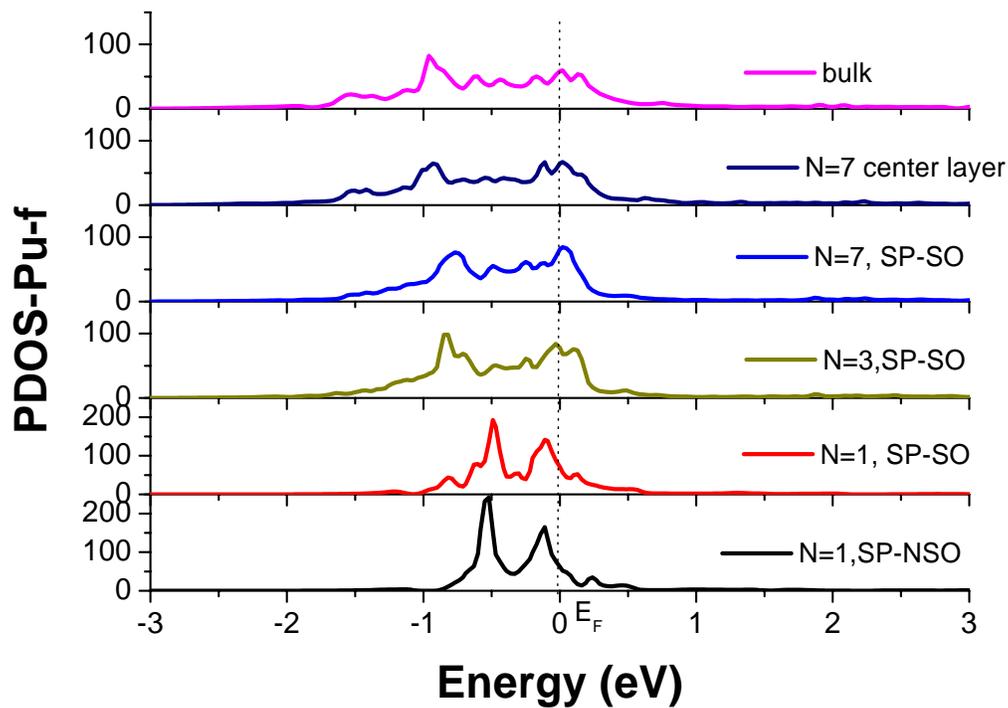

Figure 8. Density of states for Pu 5f electrons for bulk Pu and (001) N-layer slabs, where N=1, 3, 7. For the 7-layer slab, the DOS is also plotted for Pu 5f electrons in the center layer.